\documentclass{svjour3}                     
\smartqed  
\usepackage{graphicx}
\journalname{Few-Body Systems (FB20)}
\begin{document}

\title{
Electric Dipole Moments of Light Nuclei from Chiral Effective Field Theory
\thanks{Presented at the 20th International IUPAP Conference on Few-Body Problems in Physics, 20 - 25 August, 2012, Fukuoka, Japan}
}


\author{R. Higa
}


\institute{R. Higa \at
              Instituto de F\'\i sica, Universidade de S\~ao Paulo\\
              Tel.: +55-11-3091-6728\\
              Fax: +55-11-3091-6715\\
              \email{higa@if.usp.br}           
}

\date{Received: date / Accepted: date}

\maketitle

\begin{abstract}
Recent calculations of EDMs of light nuclei in the framework of chiral 
effective field theory are presented. We argue that they can be written 
in terms of the leading six low-energy constants encoding $CP$-violating 
physics. EDMs of the deuteron, triton, and helion are explicitly given in 
order to corroborate our claim. An eventual non-zero measurement of these 
EDMs can be used to disentangle the different sources and strengths of 
$CP$-violation. 
\end{abstract}

\section{Introduction}
\label{intro}

A search for new sources of $CP$-violation can be done by looking at 
electric dipole moments (EDMs) of quantum 
objects like fundamental particles, nuclei, atoms and some molecules. 
A permanent EDM of a particle violates time-reversal symmetry ($T$) which, 
according to the $CPT$ theorem, is equivalent to $CP$ violation. 
$CP$-violating phases in the CKM matrix provides an EDM to the neutron 
of the order of $|d_n|\sim 10^{-32}e\,{\rm cm}$~\cite{edm-ckm}. 
On the other hand, the existing upper limits on the neutron 
EDM~\cite{Baker06}, $|d_n|<2.9\times 10^{-26}e\,{\rm cm}$, and the proton 
EDM ($|d_p|<7.9\times 10^{-25}e\,{\rm cm}$, 
from ${}^{199}{\rm Hg}$ EDM~\cite{Griffith09}) constrain the vacuum angle 
to be $\theta<10^{-9}$. With new prospects of improving the 
sensitivity of current EDM measurements down up to two orders of 
magnitude~\cite{Baker06}, together with plans to measure EDMs of 
charged light nuclei in storage rings~\cite{edm-storing}, any non-zero 
EDM signal will either indicate new physics or set the magnitude of the 
QCD $\theta$ parameter. A method to disentangle the different 
$T$-violating ($\not\!T$) sources that can be assessed by EDMs is 
therefore called for. 
As we are going to see, the chiral symmetry of QCD can be explored to 
accomplish such task. 

\section{Chiral constraints}

Chiral effective field theory ($\chi$EFT) is nowadays a well-established 
method to deal with the non-perturbative QCD dynamics in terms of low-energy 
hadronic degrees of freedom ---see, for instance, Refs.~\cite{chptrevs}. 
SM electroweak operators can easily be incorporated via their chiral 
behavior, and here we use the same idea to deal with the leading 
$\not\!T$ interactions~\cite{thetaEFT,TVpapers}. 

\subsection{$\not\!T$ operators at the QCD scale}

The $\not\!T$ interactions can be written in terms of quark, gluon, and 
photon degrees of freedom at the QCD energies $M_{QCD}\sim$~1~GeV, well 
below its characteristic $M_{\not T}$ scale. 
The $\theta$-term and other (up to dimension-6) $\not\!T$ operators from possible extensions 
of the SM are effectively represented at the QCD scale by the following 
operators~\cite{BSMeff,TVpapers} 
\begin{eqnarray}
{\cal L}_{(\not T)}
&=&
\underbrace{ {m_{\star}}\theta\bar q\,i\gamma_5 q }_{\theta\mbox{-term}}
-\underbrace{ \frac{1}{2}\bar q({\bar d_0}
+{\bar d_3}\tau_3)\sigma^{\mu\nu}i\gamma_5\lambda^aqG^a_{\mu\nu} 
}_{\rm qCEDM}
-\underbrace{ \frac{1}{2}\bar q({d_0}+{d_3}\tau_3)
\sigma^{\mu\nu}i\gamma_5qF_{\mu\nu} }_{\rm qEDM}
\nonumber\\
&+&\underbrace{ \frac{d_{W}}{6}\varepsilon^{\mu\nu\alpha\beta}f^{abc}
G^{a}_{\mu\rho} G^{b\,\rho}_{\nu} G^{c}_{\alpha\beta} 
}_{\rm gCEDM}
+\underbrace{ \frac{1}{4}{{\rm Im}\Sigma_1}
(\bar qq\,\bar q i\gamma_5 q-\bar q\vec\tau q\cdot\bar q \vec\tau i\gamma_5 q)
}_{\rm 4q_1}
\nonumber\\
&+&\underbrace{ \frac{1}{4}{{\rm Im}\Sigma_8}
(\bar q\lambda^aq\,\bar q\lambda^a i\gamma_5 q
-\bar q\lambda^a\vec\tau q\cdot\bar q\lambda^a \vec\tau i\gamma_5 q)
}_{\rm 4q_8}\,,
\end{eqnarray}
where $m_{\star}=m_um_d/(m_u+m_d)\sim m_{\pi}^2/M_{QCD}$, with $m_{\pi}$ 
the pion mass, 
$G_{\mu\nu}$ and $F_{\mu\nu}$ are the usual gluon and photon field strengths. 
See~\cite{thetaEFT,TVpapers} for details. 
The $\theta$-term operator is the 
$4^{\underline{\rm th}}$ component of a chiral vector, Lorentz pseudoscalar 
$P=(\bar q\vec\tau q,\,\bar q\,i\gamma_5 q)$, the 
$3^{\underline{\rm rd}}$ component of it 
being an isospin-breaking term. Therefore, isospin-breaking and $\not\!T$ 
from the $\theta$-term have a close relation via chiral symmetry that can 
be further explored~\cite{thetaEFT}. 
The isoscalar and isovector terms of the quark 
chromo-EDM (qCEDM) operator, proportional to $\bar d_0$ 
and $\bar d_3$, transform as the 
$4^{\underline{\rm th}}$ and $3^{\underline{\rm rd}}$ component 
of the chiral vectors 
$
{\tilde V}=\frac{1}{2}\left(\bar q\sigma^{\mu\nu}\vec\tau\lambda^a q,
i\bar q\sigma^{\mu\nu}\gamma_5\lambda^a q\right)G^a_{\mu\nu}
$
and 
$
{\tilde W}=\frac{1}{2}\left(-i\bar q\sigma^{\mu\nu}\gamma_5
\vec\tau\lambda^a q,\bar q\sigma^{\mu\nu}\lambda^a q\right)G^a_{\mu\nu}
$,
with no useful relation to $T$-conserving operators. Analogous remarks also 
apply to the $d_0$ and $d_3$ terms in the quark EDM (qEDM) operator. 
On the other hand, the gluon chromo-EDM 
(gCEDM) and the four-quark operators ($4q_1$, $4q_8$) behave as singlets 
under the chiral group~\cite{TVpapers} ---chiral symmetry cannot disentangle 
these three interactions--- therefore they are grouped together in a 
chiral invariant (${\chi}{\rm I}$) operator. 
In terms of chiral objects, the leading $\not\!T$ Lagrangian reads 
\begin{equation}
{\cal L}^{(\not T)}={m_{\star}}\theta\,P_4-{d_0}\,{V_4}+{d_3}\,{W_3}
-{\tilde d_0}\,{\tilde V_4}+{\tilde d_3}\,{\tilde W_3}+d_w\,{\chi}{\rm I}\,.
\label{eq:Lchiop}
\end{equation}
In terms of $M_{\not T}$, $\bar m=\frac{1}{2}(m_u+m_d)$, electric 
charge $e$, and dimensionless couplings $\delta_{0,3}$, $\tilde\delta_{0,3}$, 
and $w$, the coefficients $d_{0,3}$, $\tilde d_{0,3}$, and $d_w$ are 
expected to scale as~\cite{lightEDMs} 
\begin{equation}
d_{0,3}\sim\mathcal{O}\Big(e\delta_{0,3}
\frac{\bar{m}}{M_{\not T}^{2}}\Big)\,,
\qquad
\tilde{d}_{0,3}\sim \mathcal{O}\Big(4\pi\tilde{\delta}_{0,3}
\frac{\bar{m}}{M_{\not T}^{2}}\Big)\,,
\qquad
d_{W}\sim\mathcal{O}\Big(\frac{4\pi w}{M_{\not T}^{2}}\Big)\,.
\end{equation}

\subsection{$\not\!T$ operators at the hadronic scale}

$\chi$EFT provides a systematic way of building up the chiral operators in
Eq.(\ref{eq:Lchiop}) in terms of hadronic degrees of freedom.
At LO the most relevant terms to EDMs are~\cite{lightEDMs}
\begin{eqnarray}
{\cal L}^{(\not T)}_{eff}&=&-2\bar N({\bar d_0}+{\bar d_1}\tau_3)
S^{\mu}Nv^{\nu}F_{\mu\nu}
-\frac{1}{f_{\pi}}\bar N(({\bar g_0}\vec\tau\cdot\vec\pi+{\bar g_1}\pi_3)N
\nonumber\\
&&+{\bar C_1}\bar NN\partial_{\mu}(\bar NS^{\mu}N)
+{\bar C_2}\bar N\vec\tau N\cdot\partial_{\mu}(\bar NS^{\mu}\vec\tau N)
\end{eqnarray}
where the above six low-energy constants (LECs) receive contributions
from dimensions 4 and 6 $\not\!T$ operators with distinct weights.
Given the isospin-breaking parameter $\varepsilon=(m_d-m_u)/(m_d+m_u)$
and using naive dimensional analysis (NDA) one arrives at the following
estimates~\cite{lightEDMs}:
\begin{equation}
\begin{array}{rl}
\mbox{\underline{$\theta$-term:}}&
\bar d_{0,1}\sim \theta\,e\frac{m_{\pi}^2}{M_{QCD}^3}\,,\quad
\bar g_{0}\sim \theta\frac{m_{\pi}^2}{M_{QCD}}\,,\quad
\bar g_{1}\sim \theta\,\varepsilon\frac{m_{\pi}^4}{M_{QCD}^3}\,,\quad
\\[0.2cm]
%
\mbox{\underline{qEDM:}}&
\bar d_{0,1}\sim \delta_{0,3}\,e\frac{m_{\pi}^2}{M_{QCD}M_{\not T}^2}\,,
\\[0.2cm]
%
\mbox{\underline{qCEDM:}}&
\bar g_{0}\sim (\tilde\delta_{0}\!+\!\varepsilon\tilde\delta_{3})
\frac{m_{\pi}^2M_{QCD}}{M_{\not T}^2}\,,\quad
\bar g_{1}\sim \tilde\delta_{3}\frac{m_{\pi}^2M_{QCD}}{M_{\not T}^2}\,,\quad
\bar d_{0,1}\sim e(\tilde\delta_{0}\!+\!\varepsilon\tilde\delta_{3})
\frac{m_{\pi}^2}{M_{QCD}M_{\not T}^2}\,,
\\[0.2cm]
%
\mbox{\underline{$\chi$I:}}&
\bar d_{0,1}\sim w\,e\frac{M_{QCD}}{M_{\not T}^2}\,,\quad
\bar g_{0,1}\sim w(1,\varepsilon)\frac{m_{\pi}^2M_{QCD}}{M_{\not T}^2}\,,\quad
\bar C_{1,2}\sim w\frac{M_{QCD}}{f_{\pi}^2M_{\not T}^2}\,.
\end{array}
\label{eq:scalings}
\end{equation}

\subsection{EDM of a light nucleus}

The EDM of an $A\geq 2$ nucleus receives in general two distinct 
contributions: a $\not\!T$ dipole operator $\vec D_{\not T}$ (derived from 
$\not\!T$ electromagnetic current $J^{\mu}_{\not T}$) and evaluated between 
the nucleus bra-ket states $\langle\Psi_A|$ and $|\Psi_A\rangle$, 
and a $T$-conserving dipole operator $\vec D_{T}$ (derived from 
$J^{\mu}_{T}$) evaluated between $\langle\Psi_A|$ and the corresponding 
ket-state with a $\not\!T$ admixture $|\widetilde\Psi_A\rangle$. 
The expression reads~\cite{lightEDMs} 
\begin{equation}
d_{A}=\frac{1}{\sqrt{6}}\left[
\langle\Psi_{A}||\vec{D}_{\not T}||\Psi_{A}\rangle 
+2\,\langle\Psi_{A}||\vec{D}_{T}||\widetilde{\Psi}_{A}\rangle \right]\,,
\end{equation}
where $|\Psi_A\rangle$ and $|\widetilde\Psi_A\rangle$ satisfy 
$
(E-H_0-V_{T})|\Psi_{A}\rangle=0
$ and $
(E-H_0-V_{T})|\widetilde{\Psi}_{A}\rangle=V_{\not T}|\Psi_{A}\rangle
$.
The $T$ and $\not\!T$ electromanetic currents, as well as the $\not\!T$ 
nucleon-nucleon potential $V_{\not T}$~\cite{TVpot} are derived from the 
$T$ and $\not\!T$ 
chiral effective Lagrangians. However, for ${}^2{\rm H}$, ${}^3{\rm H}$, 
and ${}^3{\rm He}$ considered in this work we use the $T$ 
wave functions from realistic phenomenological potentials. 
This hybrid approach is justified whenever the short-distance details are 
not relevant, which is partially confirmed when comparing our results with 
previous studies~\cite{oldEDMs}. 
Our results are summarized in the Tables~\ref{tab1}. 
The LECs $\bar d_0$ and $\bar d_1$ were renormalized in a way to produce 
the neutron and proton EDMs, 
$d_n=\bar d_0-\bar d_1$ and $d_p=\bar d_0+\bar d_1$. 
In general one expresses the EDMs of light nuclei in terms of the six LECs 
in different combinations, as one sees for helion and triton. 
For $N=Z$ nuclei, however, isospin selection rules makes this dependence go 
down to three, as one can verify for the deuteron case~\cite{lightEDMs}. 

\begin{table}[htb]
\begin{tabular}{||c|cccccc||}
\hline
LEC
& $\bar{d}_{0}$  & $\bar{d}_{1}$
& $(\bar{g}_{0}/{F_{\pi}})\, e\,$fm
& $(\bar{g}_{1}/{F_{\pi}})\, e\,$fm
& $({F_{\pi}}^3\bar{C}_{1})\, e\,$fm
& $({F_{\pi}}^3\bar{C}_{2})\, e\,$fm \tabularnewline
\hline
$d_{n}$  & $1$ & $-1$ & - & -  & - & -\tabularnewline
$d_{p}$  & $1$ & $1$  & - & -  & - & -\tabularnewline
$d_{^{2}{ H}}$  & $2$ & $0$  & $0.0002- 0.07\beta_1$  & $-0.19$
& - & -\tabularnewline
$d_{^{3}{ He}}$ & $0.83$ & $-0.93$  & $-0.15$  & $-0.28$
& $-0.01$ & $0.02$\tabularnewline
$d_{^{3}{ H}}$ & $0.85$ & $0.95$  & $0.15$  & $-0.28$
& $0.01$ & $-0.02$\tabularnewline
\hline
\end{tabular}
\caption{Dependence of the EDMs of the neutron, proton, deuteron, helion,
and triton on the six $\not\!T$ LECs.}
\label{tab1}
\end{table}

Although one needs six independent EDM measurements to pin down the 
six LECs, there are still some predictions that can be tested only with 
the nuclei considered here, depending on the dominance of the $\not\!T$ 
source. From the scalings of Eq.(\ref{eq:scalings}) one gets 

\begin{itemize}

\item qEDM: $d_{{}^2{\rm H}}\simeq d_n\!+\!d_p$, 
$d_{{}^3{\rm He}}\!+\!d_{{}^3{\rm H}}\simeq 0.84(d_n\!+\!d_p)$, and 
$d_{{}^3{\rm He}}\!-\!d_{{}^3{\rm H}}\simeq 0.94(d_n\!-\!d_p)$;
\vspace{2mm}

\item qCEDM: 
$d_{{}^3{\rm He}}+d_{{}^3{\rm H}}\simeq 3d_{{}^2{\rm H}}$; 
\vspace{2mm}

\item $\theta$-term: 
$d_{{}^3{\rm He}}\!+\!d_{{}^3{\rm H}}\simeq 0.84(d_n\!+\!d_p)$ 
and 
$d_{{}^3{\rm He}}\!-\!d_{{}^3{\rm H}}\neq -1.88\bar d_1
\simeq -0.94(d_n\!-\!d_p)$.

\end{itemize}
Disentangling $\chi{\rm I}$ sources is much harder to achieve since all 
LECs are involved and may appear with comparable
strength. If, for some unknown mechanism, all but $\bar d_{0,1}$ are
much more suppressed than NDA estimates, then it would be very difficult
to distinguish $\chi{\rm I}$ sources from qEDM.
The hope is to look at other light nuclei to check if additional relations 
among LECs can be retrieved.

To summarize, we calculated the contribution of different leading sources
of $CP$-violating interactions to the EDMs of deuteron, helion, and triton
using $\chi$EFT. We argue that EDMs of light nuclei can be expressed, in
general, in terms of six $\not\!T$ LECs. For $N=Z$ nuclei, isospin
selection rules probably reduce this number, as in the deuteron case.
From our expressions, exploring the distinct chiral properties and using
naive dimensional analysis, one is able to derive relations among EDMs and,
in principle, disentangle the different $\not\!T$ sources.

\begin{acknowledgements}
I would like to thank Jordy de Vries, Emmanuele Mereghetti, Rob Timmermans,
Bira van Kolck, C.-P. Liu, and Ionel Stetcu for the opportunity to
collaborate in this project. This work was partially supported by
Pr\'o-Reitoria de Pesquisa da Universidade de S\~ao Paulo and the
Dutch Stichting FOM under program 104. 
\end{acknowledgements}



\end{document}